\documentclass[12pt,epsfig]{article}
\usepackage{graphicx}
\usepackage{amsmath} 
\newcommand{\be}{\begin{equation}}
\newcommand{\ee}{\end{equation}}
\newcommand{\gsim}{\, \raisebox{-0.8ex}{$\stackrel{\textstyle >}{\sim}$ }}

\newcommand{\roughly}[1]% 
{\mathrel{\raise.4ex\hbox{$#1$\kern-.75em\lower1ex\hbox{$\sim$}}}} 

\newcommand\CL{{\cal L}}

\newcommand\wt{\tilde}

\newcommand\beq{\begin{eqnarray}} 
\newcommand\eeq{\end{eqnarray}} 
\newcommand\eqn[1]{\label{eq:#1}} 
 
\newcommand\eq[1]{eq.~(\ref{eq:#1})}

\def\Dsl{\,\raise.15ex \hbox{/}\mkern-12.8mu D} 
\newcommand\Tr{{\rm Tr\,}}

\def\fm3{fm$^{-3}$}

\begin{document}
%\begin{frontmatter}
\begin{flushright} 
%{\bf DRAFT, NOT FOR DISTRIBUTION} \\ 
INT-PUB-01-24 \\ 
MIT-CTP-3191\\
\end{flushright} 
\centerline{\LARGE Charged and Superconducting Vortices in Dense Quark Matter}
\bigskip
\bigskip
\bigskip
\bigskip
\centerline{David B. Kaplan\footnote{{\tt
        dbkaplan@phys.washington.edu}}} 
\smallskip
\centerline{\sl Institute for Nuclear Theory}
\centerline{\sl  University of Washington, 
Box 351550, Seattle, WA 98195-1550, USA }
\bigskip
\centerline{Sanjay
      Reddy\footnote{{\tt reddy@lns.mit.edu}} } 
\smallskip
\centerline{\sl  Center for Theoretical Physics, Massachusetts
  Institute of Technology}
\centerline{\sl  Cambridge, MA 02139, USA}

\begin{abstract} 
Quark matter at astrophysical densities may contain 
stable vortices due to the spontaneous breaking of hypercharge symmetry 
by kaon condensation.  We argue that these
vortices could be both charged and electrically superconducting.
Current carrying  loops (vortons) could be long lived and play
a role in the magnetic and transport properties of this matter. We
provide a scenario for vorton  formation in protoneutron stars.

\end{abstract}
\newpage
%\pacs{PACS numbers(s): 13.15.+g, 26.60.+c, 97.60.Jd}
%\end{frontmatter} 
%\documentclass[12pt]{article}
%\begin{document}

QCD with realistic quark masses possesses and exact $U(1)$ symmetry
for each flavor. Ignoring the three heaviest quarks, the relevant
symmetry is $U(1)^3 = U(1)_B\times U(1)_Y\times U(1)_{\rm em}$, where
$U(1)_B$ is baryon number, $U(1)_Y$ is hypercharge symmetry, and
$U(1)_{\rm em}$ is  the electromagnetic gauge symmetry.  These symmetries are
not explicitly broken when a chemical potential for baryon number is
turned on. However, dense quark matter exhibits color
superconductivity at sufficiently high density due to BCS-like pairing
of quarks
\cite{Barrois:1977xd,Bailin:1984bm,Alford:1998zt,Rapp:1998zu,Son:1998uk}
(for reviews of recent progress, see
\cite{Alford:2001dt,Rajagopal:2000wf,Schafer:2000et,Rischke:2000pv}).
The diquark condensate spontaneously breaks the $U(1)_B$ symmetry.
Moreover, it has recently been argued that at baryon densities
attainable inside compact stars, it is possible that either $U(1)_Y$,
$U(1)_{\rm em}$, or both are spontaneously broken by kaon condensation
\cite{Schafer:2000ew,Bedaque:2001je,Kaplan:2001qk,Schafer:2001hc}. Thus
topologically stable vortices will exist in dense matter: global
vortices due to the breaking of baryon number, and possibly global
vortices, gauged vortices, or both due to kaon condensation.  In
addition there will be vortices bounding domain walls from the
spontaneous breaking of the approximate $U(1)_A$ symmetry
\cite{Son:2000fh,Forbes:2001gj}.  In this letter we focus on global
$U(1)_Y$ vortices resulting from $K^0$ condensation.  While the
existence of such vortices has been previously mentioned in
\cite{Son:2001xd,Forbes:2001gj}, we are providing the first analysis
of  the peculiar properties of the cores of such
vortices. In this Letter we argue that the $K^0$ vortices
may in fact be electrically charged and superconducting, allowing for long
lived current loops (vortons \cite{Davis:1988jp,Davis:1989ij}).  These
vortons may affect the electromagnetic and neutrino transport
properties of dense matter.  In addition, we exploit the properties of
the phase diagram derived in \cite{Kaplan:2001qk} to provide a
plausible mechanism for the formation of these vortons.

The key to the phenomena described here is the behavior of the strange
quark in dense matter. The
$u$ and $d$ quarks are very light, but the $s$ quark is significantly
heavier, with a mass
similar to the nonperturbative QCD scale, $\Lambda_{QCD}$.  For this
reason, the strange quark does not play a big role in ordinary
matter.  However, as the baryon density of hadronic matter is
increased, introducing strange quarks can lower the energy of the
system, either in the form of hyperons, or through Bose-Einstein
condensation of anti-kaons ($K^-$) \cite{Kaplan:1986yq,Nelson:1987dg}.  The
important property of anti-kaons is that they are the lightest
particles carrying net strange quark number.

At very high matter density, one expects deconfinement to set in, and
the hadronic description of 
dense matter to give way to a quark matter description.
For three massless quark flavors at sufficiently high density, the
ground state of  matter is expected to be $SU(3)$ symmetric with equal
numbers of
$u$, $d$ and $s$ quarks;  it is also believed to exhibit
chiral symmetry breaking due to a quark bilinear  order parameter in the
color flavor locked (CFL) configuration 
\cite{Alford:1998mk}. This results in a nonet of massless Goldstone
bosons with  quantum numbers identical to the $\pi$, $K$, $\bar K$,
$\eta$ and $\eta'$ mesons found in the vacuum.  

When a strange quark
mass is turned on, this puts a strain on the system, which would like
to {\it reduce} its strange quark density relative to the $SU(3)$
symmetric CFL ground state.  For small enough
strange quark mass, the stress cannot overcome the quark pairing
energy \cite{Alford:1999pa}; however 
for sufficiently large strange quark mass, kaons will condense,
being the lightest excitation that can reduce the strange quark
content of the ground state
\cite{Bedaque:2001je}. In ref. \cite{Kaplan:2001qk} we used chiral
perturbation theory  to analyze the
resulting phase diagram 
for charge neutral matter, allowing for the presence of leptons.  We
found that at realistic densities and lepton chemical
potential $\mu_L=0$, a
$K^0$ condensate results; but  as one increases  $\mu_L$, one
passes through a second order phase transition to a mixed $K^0/K^+$
condensate\footnote{In ref. \cite{Kaplan:2001qk} we considered only
  the possibility of a homogeneous mixed $K^0/K^+$ phase; one might
  expect, however, that a heterogenous phase would be preferred.  That
  possibility is under investigation by the authors.  We thank Guy
  Moore for discussions on this subject.}, and then eventually through another second order phase
transition to a pure $K^+$ condensate.  These phases have less exact 
symmetry than than the QCD Lagrangian:  The $K^0$ condensate breaks
$U(1)_Y$ symmetry, the $K^+$ condensate breaks $U(1)_{\rm em}$
symmetry, and the mixed phase breaks both.  It follows that the three
phases will contain stable vortices --- global, gauged, or
both\footnote{The possibility of $U(1)_Y$ vortices and domain walls in
  the   $K^0$ condensate  has been recently  considered in
  refs. \cite{Son:2001xd,Forbes:2001gj}.}. 

We will focus on the global vortex found in the $K^0$ condensed phase
at zero lepton chemical potential, as is relevant for mature pulsars. 
The  starting point in the analysis of  ref. \cite{Kaplan:2001qk} is
the chiral Lagrangian at fixed charge chemical potential $\mu_Q$
describing excitations about the $SU(3)$ 
symmetric CFL vacuum, perturbing in the nonzero quark masses and electric
couplings which explicitly break the chiral $SU(3)\times SU(3)$
symmetry \footnote{Using the  low energy theory about the
false vacuum (the $SU(3)$ symmetric CFL phase) to discover the true
ground state is justified  so long as the new
phases are connected to the false vacuum by either
second order phase transitions, or first order phase transitions with
sufficiently small latent heat and barrier height.}:
\beq
\eqn{leff}
\CL &=& f_{\pi}^2 \left[\frac{1}{4}\Tr D_0 \Sigma D_0 \Sigma^\dagger  
-\frac{v^2}{4}\Tr \vec\nabla \Sigma\cdot\vec\nabla  \Sigma^\dagger \right.\cr
&& \qquad \left. + \frac{a}{2} \Tr
\tilde M\left(\Sigma + \Sigma^\dagger\right) + \frac{b}{2}\Tr Q
\Sigma Q \Sigma^\dagger\right]\ \,
\eeq
where $v$ is the in medium meson velocity and 
\beq
D_0\Sigma = \partial_0 \Sigma - i\left[
\left(\mu_Q Q + X_L\right)\Sigma -  \Sigma\left(\mu_Q
Q + X_R\right) \right]\,.
\eeq
Here $\Sigma=\exp{2i{\pi}_aT_a/f_{\pi}}$ is the $SU(3)$ matrix ($T_a$
are the $SU(3)$ generators)
describing the  meson  octet field $\pi_a$; for simplicity we have
neglected the $\eta'$. 
The meson decay constant $f_\pi$   has been computed in ref.
\cite{Son:1999cm,Son:2000tu}.
$Q$ is
the electric charge matrix ${\rm diag}(2/3,-1/3,-1/3)$,  $\tilde M =
|M| M^{-1}$ where $M$ is the
quark mass matrix  ${\rm diag}(m_u, m_d, m_s)$, and  $X_{L,R}$ are
the Bedaque-Sch\"afer terms \cite{Bedaque:2001je}:
\beq
X_L = -\frac{M M^\dagger}{2\mu}\ ,\qquad 
X_R = -\frac{ M^\dagger M}{2\mu}\ \,.
\eeq 
  From this
Lagrangian one derives the kaon masses and chemical potentials
\beq
%m^2_{\pi^-} &=& a (m_u + m_d)m_s + b \cr 
m^2_{K^+} = a (m_u + m_s)m_d + b\ ,\qquad
m^2_{K^0} = a (m_d + m_s)m_u \,.
\eqn{masses}
\eeq
\beq
%\tilde \mu_{\pi^+} = \mu_Q + \frac{m_d^2 - m_u^2}{2\mu}\ ,\quad
\tilde \mu_{K^+} = \mu_Q + \frac{m_s^2 - m_u^2}{2\mu}\ , \quad
\tilde \mu_{K^0} = \frac{m_s^2 - m_d^2}{2\mu}\ .
\eeq

One can show that the free energy derived from this Lagrangian is
minimized for the classical vacuum configuration $\Sigma(x,t) =
e^{-i\tilde \mu t} \Sigma(x)  e^{i\tilde \mu t}$, where 
\beq
\tilde \mu = \mu_Q Q  + X\ ,\qquad X=X_L=X_R=-\frac{M^2}{2\mu}\ .
\eeq
The time-independent field $\Sigma(x)$ minimizes the free energy
\beq
\Omega &=& \frac{f_{\pi}^2}{4}\left[\Tr [\tilde\mu,\Sigma][
  \tilde\mu,\Sigma^\dagger]+v^2\,\Tr \vec\nabla \Sigma\cdot\vec\nabla
  \Sigma^\dagger \right.\cr &&\qquad\left.
- 2a\, 
 \Tr \wt M (\Sigma  + \Sigma^\dagger-2) - b\,\Tr [Q,
  \Sigma][ Q ,\Sigma^\dagger]\right] \ , 
\eqn{omega}
\eeq
which is equivalent to requiring $\Sigma(x)$ to obey the equation of
motion
\beq
\left[\frac{1}{2}\vec\nabla\cdot\left(\Sigma^\dagger \vec\nabla
    \Sigma\right)+ \tilde\mu\Sigma^\dagger \tilde\mu \Sigma - a\,
  \tilde M\Sigma - b\,  Q
  \Sigma^\dagger 
Q \Sigma\right] -h.c.=0 \ .
\eqn{sigmin}
\eeq

In ref. \cite{Kaplan:2001qk} we found that for $\mu_Q=0$ the minimum
free energy
configuration was described by a spatially constant $K^0$ condensate with 
\beq
\cos \phi_0 =
\frac{M_{K^0}^2}{\mu_{K^0}^2} \ ,\qquad \phi_0\equiv \frac{\langle K^0
  \rangle}{\sqrt{2} f_\pi}\ ,
\eqn{kz}
\eeq
for $\mu_{K^0}> {M_{K^0}}$.  This latter condition seems to be  met for
reasonable parameters one might encounter in the core of a compact
star. An assumption behind this conclusion is that instanton effects
are suppressed;  these would induce a $\Tr M\Sigma$ operator
\cite{Alford:1998mk}, enhancing the kaon mass.

Since a nonzero $K^0$ field spontaneously breaks the exact $U(1)_Y$
symmetry of QCD, it follows that there must be topologically stable
vortex solutions as well of the form 
\beq
 \frac{K^0(r,\theta)}{\sqrt{2} f_\pi} =
\phi(r)e^{i\theta}\ , \qquad {\rm with\ }\quad \phi(0)=0\ ,\quad
\phi(\infty)=\phi_0\ . 
\eeq
The profile $\phi(r)$ can be computed numerically from \eq{sigmin},
with the results for several values of vacuum condensate $\phi_0$
shown in Fig.~1 as solid curves. In the absence of weak interactions,
these  vortices arise from the spontaneous breaking of an exact
$U(1)_Y$  global symmetry, and so  aligned  vortex-antivortex pairs exert
an attractive force that varies logarithmically with the separation,
and an isolated vortex has a logarithmically divergent energy per unit
length.  Including the weak interactions which explicitly break
$U(1)_Y$ symmetry, leads to domain wall formation and a linear
potential between vortex and antivortex pairs \cite{Son:2001xd}. The
scale associated with transition from logarithmic to linear behavior in the
potential has been estimated in ref. \cite{Son:2001xd}  to be $\sim 10^3$~fm.

%%%%%%%%%%%%%%%%%%%%%%%%%%%%%%%%%%%%%%%%%%%
\begin{figure}[t]
\begin{center}
\includegraphics[width=.85\textwidth,angle=-90]{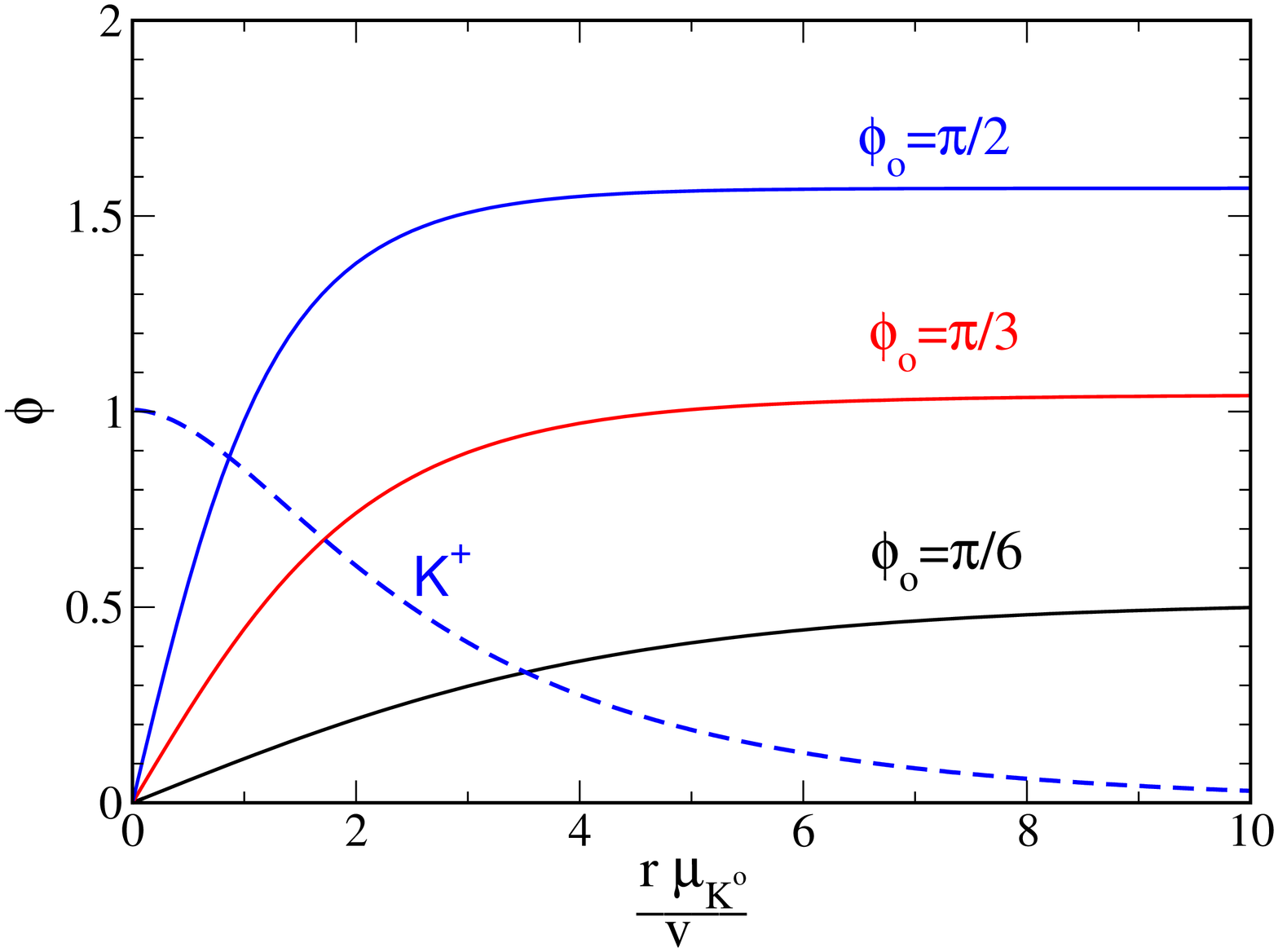}
\end{center}
\caption{The profile $\phi(r)$ of the global vortex present in the
  $K^0$ condensed phase of dense quark matter, plotted versus radial
  distance $r v/\mu_K^0$ for different values of the
  asymptotic condensate  $\phi_0$ (solid lines). The angle $\phi$ is related to
  the $K^0$ condensate amplitude by $\phi=\frac{|\langle
    K^0\rangle|}{\sqrt{2} f_\pi}$, and $\cos\phi_0 =
  M_{K^0}^2/\mu_{K^0}^2$ \cite{Kaplan:2001qk}. Shown in dashed line is
  the profile of the unstable $K^+$ mode which condenses in the vortex
  core for the case $\phi_0=\pi/2$. 
}
\label{profile}
\end{figure}
%%%%%%%%%%%%%%%%%%%%%%%%%%%%%%%%%%%%
The $K^0$ condensate necessarily vanishes in the core of the vortex,
and hence locally there is an equal density of $u$,
$d$ and $s$ strange quarks, in spite of the strange quark
being massive.  This suggests that the $K^0$ vortex might be unstable
to the condensation of $K^+$ mesons in the core region, which would
serve to reduce the strange quark number.

A crude estimate is useful for determining whether or not $K^+$ mesons
might condense in the vortex core. Noting that the core size of
the $K^0$ vortices shown in Fig.~1 is $r_0\sim c \frac{v }{\mu_{K^0}}$
where $c=O(1)$, the energy cost of creating a single $K^+$ localized  in the core is $\delta E\sim \sqrt{M_{K^+}^2 + 
  \mu_{K^0}^2/c^2}$. An instability will set in if $\delta E < 
\mu_{K^+}$.  For $\mu_Q=0$, we have $\mu_K^+\simeq \mu_{K^0}\simeq
m_s^2/(2 \mu)$. 
Furthermore, for densities relevant for compact stars and for a
superconducting gap  $\Delta \ll m_s$, one finds $M_K^+\ll \mu_{K^0}$.
Therefore, at least for $\Delta$ not too large, it appears that the
existence of an instability depends primarily on the core size, parametrized by
$c$.  

 Ideally one would like to
solve the coupled nonlinear differential equations \eq{sigmin} to find
such a $K^+$ condensate. Instead we have investigated a simpler
problem: for the case of $\phi_0=\pi/2$  (corresponding to
$M_{K^0}\ll \mu_{K^0}$) we have linearized the equation of motion for
$K^+$ excitations about the unperturbed $K^0$ vortex background
displayed in Fig.~1.  We discovered that there indeed exists an
unstable mode, whose spatial profile is indicated in Fig.~1 by the
dashed line.  The existence of this instability for $\phi_0=\pi/2$
implies that at least for some range of $\phi_0$, a $K^0$ vortex can
lower its free energy by having a $K^+$ condensate form in its core.

The existence of a $K^+$ condensate in the core of the $K^0$ vortices
could be of phenomenological interest, as it would mean that vortices
would be both electrically charged (neutralized by electrons) and
electrically superconducting, capable of carrying persistent
currents\footnote{The system described here is similar to the scalar
  model presented 
  in ref. \cite{Witten:1985eb} as an example of  cosmological
  superconducting strings.}. The current would take the form of an
angular dependent 
phase in the $K^+$ condensate as one circumnavigates the $K^0$
loop. In the vacuum, a loop of global vortex would be expected to
shrink and disappear radiating mesons.  However, a loop of vortex in the
presence of a Bose-Einstein condensate is kept from shrinking away by
the velocity dependent Magnus force\footnote{We thank D. T. Son for
  bringing the Magnus force to our attention.}; this behavior will
persist even with the attractive linear potential due to the
electroweak domain walls.

These loops, called vortons \cite{Davis:1988jp,Davis:1989ij},   are
further stabilized if they carry electric currents. 
Since the current density carried by the $K^+$ core is $K^*
i\overset{\leftrightarrow}{\nabla} K$, while the kinetic energy
density is $|\nabla K|^2$, an estimate of the energy of a vorton of
radius $R$ carrying a current 
$J$ yields  $E\sim 2\pi( R \lambda + R^2 \sigma +
c J^2/R)$, where $\lambda$ is the energy per unit length of the $K^0$
vortex (including the charge screening electrons), $\sigma$ is the
energy per unit area of the electroweak domain wall stretched like a
soap bubble across the loop, and $c$ is a 
dimensionless number expected to be $\gsim O(1)$. 
 It follows that static
vortons carrying classical currents are at least stable against
shrinking, although we have not investigated the possibility of shape
instabilities, the formation of cusps and damping due to
electromagnetic radiation \cite{Blanco-Pillado:2001wq}.

There is a plausible mechanism for the formation of vortons during the
creation of the protoneutron star.  Initially, the protoneutron star
is expected to have a sizable lepton number chemical potential $\mu_L$ due
to neutrino trapping.  As discussed in ref. \cite{Kaplan:2001qk}, 
there is a broad range of values of $\mu_L$ for which the ground state
of superconducting quark matter spontaneously breaks the $U(1)_Y\times
U(1)_{\rm em}$ symmetry of QCD down to nothing via the mixed
condensate of
 $K^0$ and $K^+$ mesons.  In such a phase, both
global $K^0$ vortices and gauged $K^+$ vortices are possible, and
would be expected to form during the stellar core
collapse.  In particular, one may expect configurations to form where
a $K^0$ vortex loop encloses net magnetic flux, carried by the  $K^+$
vortices.  As 
the neutrinos diffuse from the core, $\mu_L$ decreases, and as implied
by  the phase diagram in ref. \cite{Kaplan:2001qk}, matter goes through
a second order phase transition to the pure $K^0$ condensed phase.
When this occurs, the $K^+$ vortex will dissolve, and the bulk matter
ceases to be electrically superconducting, except in the core of the
$K^0$ vortex. Thus the $K^0$ vortex
loop will persist with a   trapped magnetic dipole field, 
and a  concomitant current flowing in its core --- a vorton.

The system we have described is quite complex, and more work
will be required to show that the scenario described above could
lead to a macroscopic density of charge and magnetic field carrying
vortons within a young compact star interior.  Since such a
possibility obviously impacts bulk properties of the young star, such
as its neutrino opacity and magnetic properties, we consider further
work on this subject warranted.

{\it Acknowledgements:}
  We wish to thank P. Bedaque, G. Moore,  K. Rajagopal, and D. T. Son for useful
  comments on an early version of this manuscript.  This work was
  supported by DOE grants DE-FG03-00-ER-41132 and DF-FC02-94ER40818.

\bibliography{vortex}

\begin{thebibliography}{10}

\bibitem{Barrois:1977xd}
B.~C. Barrois,
\newblock Nucl. Phys. {\bf B129}, 390 (1977).
%%CITATION = NUPHA,B129,390;%%

\bibitem{Bailin:1984bm}
D.~Bailin and A.~Love,
\newblock Phys. Rept. {\bf 107}, 325 (1984).
%%CITATION = PRPLC,107,325;%%

\bibitem{Alford:1998zt}
M.~G. Alford, K.~Rajagopal, and F.~Wilczek,
\newblock Phys. Lett. {\bf B422}, 247 (1998), hep-ph/9711395.
%%CITATION = HEP-PH 9711395;%%

\bibitem{Rapp:1998zu}
R.~Rapp, T.~Schafer, E.~V. Shuryak, and M.~Velkovsky,
\newblock Phys. Rev. Lett. {\bf 81}, 53 (1998), hep-ph/9711396.
%%CITATION = HEP-PH 9711396;%%

\bibitem{Son:1998uk}
D.~T. Son,
\newblock Phys. Rev. {\bf D59}, 094019 (1999), hep-ph/9812287.
%%CITATION = HEP-PH 9812287;%%

\bibitem{Alford:2001dt}
M.~G. Alford,
\newblock Ann. Rev. Nucl. Part. Sci. {\bf 51}, 131 (2001), hep-ph/0102047.
%%CITATION = HEP-PH 0102047;%%

\bibitem{Rajagopal:2000wf}
K.~Rajagopal and F.~Wilczek,
\newblock (2000), hep-ph/0011333.
%%CITATION = HEP-PH 0011333;%%

\bibitem{Schafer:2000et}
T.~Schafer and E.~V. Shuryak,
\newblock Lect. Notes Phys. {\bf 578}, 203 (2001), nucl-th/0010049.
%%CITATION = NUCL-TH 0010049;%%

\bibitem{Rischke:2000pv}
D.~H. Rischke and R.~D. Pisarski,
\newblock (2000), nucl-th/0004016.
%%CITATION = NUCL-TH 0004016;%%

\bibitem{Schafer:2000ew}
T.~Schafer,
\newblock Phys. Rev. Lett. {\bf 85}, 5531 (2000), nucl-th/0007021.
%%CITATION = NUCL-TH 0007021;%%

\bibitem{Bedaque:2001je}
P.~F. Bedaque and T.~Schafer,
\newblock Nucl. Phys. {\bf A697}, 802 (2002), hep-ph/0105150.
%%CITATION = HEP-PH 0105150;%%

\bibitem{Kaplan:2001qk}
D.~B. Kaplan and S.~Reddy,
\newblock Phys. Rev. {\bf D65}, 054042 (2002), hep-ph/0107265.
%%CITATION = HEP-PH 0107265;%%

\bibitem{Schafer:2001hc}
T.~Schafer,
\newblock (2001), nucl-th/0109029.
%%CITATION = NUCL-TH 0109029;%%

\bibitem{Son:2000fh}
D.~T. Son, M.~A. Stephanov, and A.~R. Zhitnitsky,
\newblock Phys. Rev. Lett. {\bf 86}, 3955 (2001), hep-ph/0012041.
%%CITATION = HEP-PH 0012041;%%

\bibitem{Forbes:2001gj}
M.~M. Forbes and A.~R. Zhitnitsky,
\newblock Phys. Rev. {\bf D65}, 085009 (2002), hep-ph/0109173.
%%CITATION = HEP-PH 0109173;%%

\bibitem{Son:2001xd}
D.~T. Son,
\newblock (2001), hep-ph/0108260.
%%CITATION = HEP-PH 0108260;%%

\bibitem{Davis:1988jp}
R.~L. Davis and E.~P.~S. Shellard,
\newblock Phys. Lett. {\bf B207}, 404 (1988).
%%CITATION = PHLTA,B207,404;%%

\bibitem{Davis:1989ij}
R.~L. Davis and E.~P.~S. Shellard,
\newblock Nucl. Phys. {\bf B323}, 209 (1989).
%%CITATION = NUPHA,B323,209;%%

\bibitem{Kaplan:1986yq}
D.~B. Kaplan and A.~E. Nelson,
\newblock Phys. Lett. {\bf B175}, 57 (1986).
%%CITATION = PHLTA,B175,57;%%

\bibitem{Nelson:1987dg}
A.~E. Nelson and D.~B. Kaplan,
\newblock Phys. Lett. {\bf B192}, 193 (1987).
%%CITATION = PHLTA,B192,193;%%

\bibitem{Alford:1998mk}
M.~G. Alford, K.~Rajagopal, and F.~Wilczek,
\newblock Nucl. Phys. {\bf B537}, 443 (1999), hep-ph/9804403.
%%CITATION = HEP-PH 9804403;%%

\bibitem{Alford:1999pa}
M.~G. Alford, J.~Berges, and K.~Rajagopal,
\newblock Nucl. Phys. {\bf B558}, 219 (1999), hep-ph/9903502.
%%CITATION = HEP-PH 9903502;%%

\bibitem{Son:1999cm}
D.~T. Son and M.~A. Stephanov,
\newblock Phys. Rev. {\bf D61}, 074012 (2000), hep-ph/9910491.
%%CITATION = HEP-PH 9910491;%%

\bibitem{Son:2000tu}
D.~T. Son and M.~A. Stephanov,
\newblock Phys. Rev. {\bf D62}, 059902 (2000), hep-ph/0004095.
%%CITATION = HEP-PH 0004095;%%

\bibitem{Witten:1985eb}
E.~Witten,
\newblock Nucl. Phys. {\bf B249}, 557 (1985).
%%CITATION = NUPHA,B249,557;%%

\bibitem{Blanco-Pillado:2001wq}
J.~J. Blanco-Pillado, K.~D. Olum, and A.~Vilenkin,
\newblock Phys. Rev. {\bf D63}, 103513 (2001), astro-ph/0004410.
%%CITATION = ASTRO-PH 0004410;%%

\end{thebibliography}
\bibliographystyle{h-physrev3.bst} 

\end{document}